\documentclass[11pt]{article}
\usepackage{latexsym}
\usepackage{graphicx}

\newcommand{\phib}{\ensuremath{\overline{\phi}}}

\newcommand{\beq}{\begin{equation}}
\newcommand{\eeq}{\end{equation}}
\newcommand{\DK}{Dirac-K\"{a}hler }

\begin{document}

\begin{center}
{\bf\Large Notes on (twisted) lattice supersymmetry}\\
Simon Catterall\\
Department of Physics, Syracuse University,\\
Syracuse, NY 13244, USA\\
email: smc@physics.syr.edu\\        
\end{center}

\begin{abstract}
We describe a new approach to the problem of putting supersymmetric theories
on the lattice. The basic idea is to discretize a {\it twisted} formulation
of the supersymmetric theory. For certain theories with extended
supersymmetry these twisted formulations contain only integer spin
fields. The twisting exposes a scalar nilpotent
supercharge which generates an exact lattice symmetry.  We gives examples
from quantum mechanics, sigma models and Yang-Mills theories.
\end{abstract}

\begin{center}
\Large Introduction
\end{center}
The difficulties of discretizing supersymmetric theories are well known.
Generic naive discretizations of continuum supersymmetric theories
do not preserve supersymmetry. Quantum corrections then generate a large
number of relevant supersymmetry violating interactions whose
couplings must be tuned to zero as the lattice spacing is reduced.
This is both unnatural and in many cases (especially for
models with extended supersymmetry) prohibitively difficult.
Various attempts have been made over the last twenty five years to
overcome these problems see \cite{old} and \cite{recent} and the recent
reviews \cite{feo,kap}. However most of this work was confined to
low dimensional models or Hamiltonian formulations.

Quite recently a series of new approaches have been developed which share
the common feature of preserving a sub-algebra of the full supersymmetry
algebra {\it exactly} at finite lattice spacing\footnote{Very recently a
lattice construction of ${\cal N}=2$ super Yang-Mills in $D=2$ has been proposed which
preserves {\it all} the supercharges \cite{n=2noboru}}
\cite{qm,top,kap1,kap2}. The hope is that this
exact symmetry will protect the lattice theory against at least some
of these dangerous radiative corrections and thus reduce fine
tuning. 

Theories where these ideas may be applied all possess
{\it extended} supersymmetry. While these theories are
not of immediate phenomenological interest they exhibit fascinating
connections to string and gravitational theories as exhibited
by the well known correspondence between ${\cal N}=4$ super Yang-Mills
in four dimensions
theory and type IIB string theory on AdS space. Actually, the latter
forms perhaps the best known example of a more general conjectured
duality between $p+1$-dimensional super Yang-Mills and black p-brane
solutions in supergravity. 

This review will concentrate on just one of these new approaches to 
formulating lattice supersymmetry -- discretization of a
twisted version of the supersymmetric theory \cite{top,twist,noboru1}. The 
construction applies only to cases
where the number of continuum supercharges is a multiple of
$2^D$ in $D$ dimensions. We first start with a toy model, supersymmetric
quantum mechanics realized as a $(0+1)$ dimensional field theory
and show how to realize an exact nilpotent lattice
supersymmetry in that model. We then go on to show how to lift this
model to two dimensions to construct a lattice action for the two
dimensional sigma model which retains an
exact supersymmetry at finite lattice spacing. Numerical
results deriving from full dynamical fermion simulations and
confirming
exact lattice supersymmetry in these models are presented. The general
twisting procedure is then described in two and four dimensions and
the twisted actions of ${\cal N}=2$ and ${\cal N}=4$ super Yang-Mills
in two and four dimensions are written down. The discretization of
these gauge systems is then described in some detail.
\section{Toy model}
Consider a toy model consisting of a set of 
commuting fields $\phi(t)$ and $B(t)$ depending on a single 
continuous (Euclidean)
time parameter $t$, together
with anticommuting fields $\chi(t)$ and $\psi(t)$. Let us also
postulate the 
fermionic symmetry 
\begin{eqnarray}
Q\phi&=&\psi\nonumber\\
Q\psi&=&0\nonumber\\
Q\chi&=&B\nonumber\\
QB&=&0
\label{1dQ}
\end{eqnarray}
Notice that this symmetry is nilpotent off-shell
and is reminiscent of a BRST symmetry. Using this structure we
can write down a 
action which resembles a gauge fixing term
\beq S=Q\int dt \chi\left(N(\phi)+\frac{1}{2}B\right)\eeq
Carrying out the variation we find
\beq S=\int dt\left(
BN+\frac{1}{2}B^2+\chi(\frac{\partial N}{\partial \phi})\psi\right)\eeq
After integrating over $B$ we are led to the on-shell action
\beq S=\int dt \frac{1}{2}N(\phi)^2+\chi\frac{\partial N}{\partial \phi}\psi\eeq
To recover a physical theory from this
construction it is necessary to choose
a specific function $N(\phi)$. If we choose $N(\phi)=\partial_t\phi+P(\phi)$
our action can be recognized as nothing more than 
Witten's supersymmetric quantum mechanics \cite{wsusy}.
\beq N^2\to (\partial\phi+P^2(\phi))^2\eeq
\beq \frac{\partial N}{\partial\phi}\to (\partial+P^\prime(\phi))\eeq
Notice the presence of cross terms in the bosonic action which in the
continuum are total derivatives and hence can be neglected. When we
latticize the theory their presence will be
necessary to ensure exact supersymmetry.
This correspondence requires that we associate the ``ghost'' and ``anti-ghost''
field $\psi$ and $\chi$ with physical fermion fields. This
relationship between twisted and conventional fermion fields
will become more complicated in higher dimensions. Unlike usual
BRST gauge fixing we must however {\it not} impose the physical
state conditions $Q|{\rm phys}>=0$.

\begin{figure}
\begin{center}
\resizebox{5in}{!}{\includegraphics
{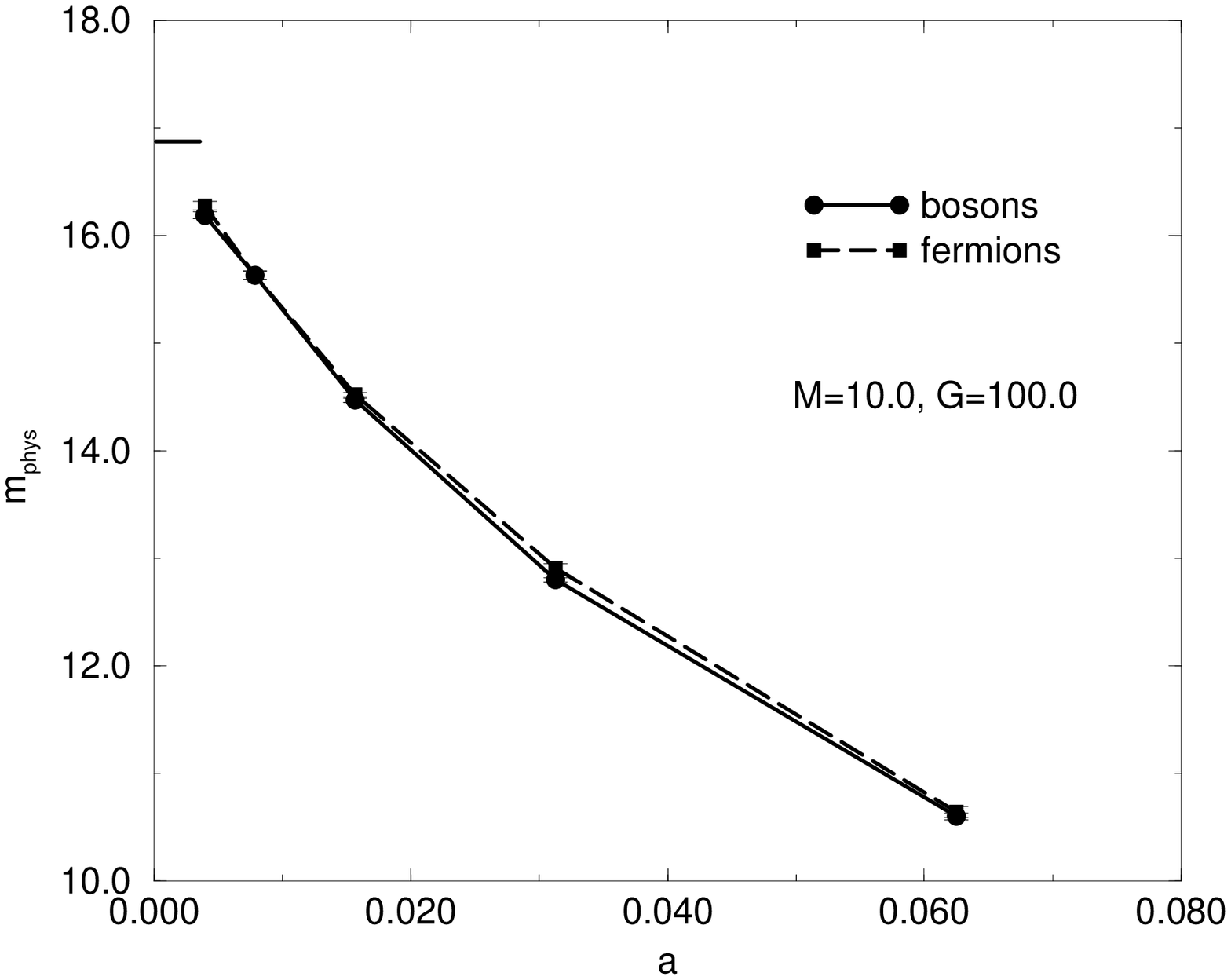}}
\end{center}
\label{figqm}
\end{figure}

Notice that invariance
of this action under $Q$ depends only on its nilpotent property -- 
not the form of the function $N(\phi)$. Indeed the twisted
supersymmetry transformation eqn.~\ref{1dQ} involves no derivatives
in time and hence can be trivially transferred to the lattice. The
resulting lattice action is 
\beq S_L=Q\sum \chi_t\left(D^+_{tt^\prime}\phi_{t^\prime}+P(\phi_t)+B_t\right)\eeq
where
\beq\Delta_{tt^\prime}^+\phi_{t^\prime}=\left(
\delta_{t^\prime,t+a}-\delta_{t^\prime,t}\right)\phi_{t^\prime}
\eeq
Notice that the use of a forward difference operator 
ensures no bosonic doublers. Exact SUSY then implies no
fermion doubles. We will see later that the use of
forward and backward difference operators is natural in higher
dimensions also and follows from regarding the fermion fields as
components of a \DK field. Well-defined discretizations of
the \DK equation necessarily introduce such operators.
Of course supersymmetric quantum
mechanics possesses two supercharges. In this formulation this
second symmetry can be gotten by exchanging $\chi\to\psi$.
Notice that this second symmetry is classically broken by a
term $O(a^2)$. However, an absence of
divergences ensures this symmetry is automatically
restored without fine tuning as $a\to
0$ \cite{qmgiedt}. In figure~\ref{figqm}
we show some numerical results for the boson and fermion
massgaps deriving from a dynamical fermion
simulation of this model using $P=m\phi+g\phi^3$ and $m=10$, $g=100$
and lattice sizes $L=16$, $L=32$, $L=64$, $L=128$ and $L=256$.
Even at the largest
lattice spacings a clear boson/fermion degeneracy can be seen. 
Furthermore, the lattice massgap appears to flow to
the correct continuum value (calculated using Hamiltonian techniques) without
fine tuning. 
We have also tested a number of Ward identities following from this
symmetry. The simplest of these is the expectation of the
bosonic action which turns out to be
\beq <S_B>=\frac{N_{\rm dof}}{2\beta}\eeq
For results on this quantity and other simple Ward
identities we refer the reader to \cite{qm,qmgiedt} in which extensive
numerical results are provided to support 
the existence of exact lattice supersymmetry and the claim  that
no fine tuning is needed in this model to take the continuum limit. 
\section{Relation to topological quantum field theory}
This BRST invariance reflects an underlying
local shift symmetry \cite{rev}. Consider a
model with a finite number of fields $\phi_i$ and
$S_{\rm cl}(\phi)=0$. This is trivially invariant
under the topological symmetry
\beq
\phi_t\to \phi_t+\epsilon_t
\eeq
To quantize this theory requires picking a gauge. Choosing
$N_t(\phi)=D^+_{tt^\prime}\phi_t^\prime+
P^\prime_t(\phi)=0$.
leads to
\beq
Z=\int \prod d\phi_t \delta(N_t) {\rm det}\left(\frac{\partial N_t}{\partial
\phi_{t^\prime}}\right)
\eeq
If we represent determinant using anticommuting ghosts and introduce a
multiplier field for the $\delta$-function we
recover our SUSY model in Landau gauge!
Notice though that this topological theory requires a projection to states
annihilated by $Q$. This is equivalent to a projection to the
vacuum state and {\it is not} what we do here. Here, the twisted or
topological field theory form is simply to be viewed as a change of
variables in the underlying supersymmetric theory. In flat space the
regular and twisted formulations are completely equivalent.
\section{Sigma model}
A possible generalization of this quantum mechanical model consists of
equipping the scalar fields with an additional index $\phi\to\phi^i$ and
regarding these fields as 
coordinates on some non-trivial target
space with metric
$g_{ij}(\phi)$ \cite{sigma}.
The appropriate nilpotent symmetry is now
\begin{eqnarray}
\label{aa}
Q\phi^i&=&\psi^i \nonumber \\
Q\psi^i&=&0 \nonumber \\
Q\chi_i&=&\left(B_i-\chi_j\Gamma^j_{ik}\psi^k\right) \nonumber \\
QB_i&=&\left(B_j\Gamma^j_{ik}\psi^k-\frac{1}{2}\chi_j
R^j_{ilk}\psi^l\psi^k\right)\nonumber
\end{eqnarray}
and the appropriate gauge fermion looks like 
\beq
\Psi=\int_\sigma \eta_i\left(N^i\left(\phi\right)-
\frac{1}{2}g^{ij}B_j\right)
\eeq
with action $S=\beta Q\Psi$.
Carrying out the variation and integrating out $B$ as before we find a twisted
form of the usual supersymmetric sigma model action
\beq
S=\alpha\int_\sigma
\left(\frac{1}{2}g_{ij}N^iN^j-\chi_i \nabla_kN^i\psi^k+
\frac{1}{4}R_{jlmk}\chi^j\chi^l\psi^m\psi^k\right)
\eeq
which is then invariant under the scalar supersymmetry
\begin{eqnarray}
Q\phi^i&=&\psi^i\nonumber\\
Q\psi^i&=&0\nonumber\\
Q\chi_i&=&\left(g_{ij}N^j-\chi_j\Gamma^j_{ik}\psi^k\right)\nonumber
\end{eqnarray}
We still must specify the ``gauge fixing'' function $N^i$. For a one
dimensional base space with coordinate $\sigma$ we can just take
$N^i=\frac{d\phi^i}{d\sigma}$ and the resulting action is
\begin{equation}
S=\beta\int d\sigma
\left(\frac{1}{2}g_{ij}\frac{d\phi^i}{d\sigma}\frac{d\phi^j}{d\sigma}
-\chi_i \frac{D}{D\sigma} \psi^i+
\frac{1}{4}R_{jlmk}\chi^j\chi^l\psi^m\psi^k\right)
\end{equation}
where the covariant derivative is the pullback of its target space cousin
\begin{equation}
\frac{D}{D\sigma}\psi^i=\frac{d}{d\sigma}\psi^i+\Gamma^i_{kj}\frac{d\phi^k}{d\sigma}\psi^j
\end{equation}
Discretization of this action is just the same as for quantum mechanics and
proceeds by replacing a continuum derivative by a forward difference operator.

The situation becomes more interesting when we take the base space to be
two dimensional. The natural gauge fixing term now becomes
\begin{equation}
N^{i\alpha}=\partial^\alpha\phi^i
\end{equation}
and implies that the anti-ghost 
$\chi$ and multiplier $B$ also
acquire an additional base space vector index.
Actually this choice will not
do. It is clear that if we are to arrive at a supersymmetric
model the number of degrees of freedom carried by the anti-ghost must
match that of the ghost field (in the end they will turn out to correspond to 
different chiral components of the physical fermions).
Thus we must require the anti-ghost
$\chi^{i\alpha}$ and
multiplier field $B^{i\alpha}$ satisfy some condition which halves their
number of degrees of freedom \cite{rev}. The natural way to do this is to introduce
projection operators $P^{\left(-\right)}$ and $P^{\left(+\right)}$ and
require that $\chi^{i\alpha}$ and $B^{i\alpha}$ satisfy 
certain self-duality
conditions
\begin{eqnarray}
P^{\left(-\right)}\chi&=&0\nonumber\\
P^{\left(+\right)}\chi&=&\chi
\label{self-dual}
\end{eqnarray}
One choice for these projectors is
\begin{equation}
{P^{i\alpha}_{j\beta}}^{\left(\pm\right)}=
\frac{1}{2}\left(\delta^i_j\delta^\alpha_\beta\pm
J^i_j\epsilon^\alpha_\beta\right)
\end{equation}
Here, $J^i_j$ must be a globally defined tensor field
on the
target space which squares to minus the identity and $\epsilon^\alpha_\beta$ is the usual antisymmetric
matrix with constant coefficients. Manifolds
possessing such a structure are 
called {\it almost complex} and have
even dimension. At this point we must be careful to make sure that
the BRST transformations we introduced earlier  
are compatible with
these self-duality conditions.
This constraint forces the almost complex structure to be
covariantly constant $\nabla_kJ^i_j=0$ and the
manifold is termed K\"{a}hler.
The final action in complex coordinates takes the form
\begin{eqnarray}
S&=&\beta\int d^2\sigma\left(
2h^{+-}g_{I\overline{J}}\partial_+\phi^I
\partial_-\phi^{\overline{J}}\right.\nonumber\\
&-&\left.h^{+-}g_{I\overline{J}}\chi^I_+D_-\psi^{\overline{J}}-
h^{+-}g_{\overline{I}J}\chi^{\overline{I}}_-D_+\psi^J+
\frac{1}{2}h^{+-}R_{I\overline{I} J\overline{J}}
\chi^I_+\chi^{\overline{I}}_-\psi^J\psi^{\overline{J}}\right)
\end{eqnarray}
It should be clear by inspection that this model is indeed
the ${\cal N}=2$ supersymmetric sigma model with $\chi_+^I$ and $\psi^I$
corresponding to the Weyl components of a Dirac spinor $\lambda^I$
in chiral basis \cite{witten2,alvarez,zumino}.

To date discretization of this action has been effected by replacing the
continuum derivative by a {\it symmetric} difference operator and adding
an additional Wilson term in the form of a holomorphic Killing vector
to preserve the $Q$-exactness of the action. For details we refer
to \cite{sigma}. It should also be possible to proceed 
by rewriting the fermionic
action in \DK form and utilizing the same discretization prescription we
will advocate later for Yang-Mills theories. This has yet to be done.

One other interesting limit occurs for these two dimensional theories
if I take a flat two dimensional target space. In this case
I can deform the model {\it without losing the
$Q$-exactness of the action} by addition of a holomorphic potential and obtain
the two dimensional complex Wess-Zumino model. We refer the
interested reader to \cite{wz} for details and extensive numerical
simulations.
 
To summarize we have shown that it is possible to find lattice
formulations of one and two dimensional supersymmetric theories with
extended (${\cal N}=2$) supersymmetry which
are exactly invariant under a single scalar fermionic symmetry.
Furthermore this fermionic symmetry corresponds
to a particular combination of the usual supercharges
and emerges naturally in the context of twisted or topological
field theory formulations of the supersymmetric theory. These
twisted formulation naturally contain scalar and vector fermions.

There are two problems in what we have said so far; firstly
we have not given a general method for constructing the twisted
variables in terms of the usual fields. This we will rectify in the
next section. Second and more important is the fact that
we have not considered theories with a gauge symmetry. Twisted
formulations of gauge theories exist in the continuum (indeed the
very first topological field theory constructed by Witten
corresponded to twisted ${\cal N}=2$ super Yang-Mills in four
dimensions) but the twin
requirements of exact gauge symmetry and exact supersymmetry render
a simple translation of the continuum constructions to the 
lattice problematic. However, progress was made 
when in \cite{sugino}, Sugino managed
to generalize the twisted supersymmetry transformations to the
lattice. Unfortunately, his construction generically yields additional
states in the lattice theory with no counterpart in the continuum.
In low dimensions it is possible to circumvent these problems by
careful choice of the gauge fermion but this approach fails in
four dimensions.
In \cite{2dpaper} we proposed an alternative discretization of
these twisted models which does not suffer from these problems. We will
spend the last part of this article reviewing this approach which we will
see includes the interesting case of ${\cal N}=4$ super Yang-Mills.
\section{Twisting as a change of variables}
The twist required to expose a scalar supercharge in ${\cal N}=2$ 
and ${\cal N}=4$ super Yang-Mills theory in $D=2$ and $D=4$ dimensions
respectively is gotten by replacing the usual rotation group
$SO(D)$ by the diagonal subgroup \cite{noboru2,noboru4,2dpaper,4dpaper}
\beq
SO(D)^\prime={\rm diagonal \;subgroup}(SO(D)\times SO(D)_R)\eeq
where the second factor reflects the additional R-symmetry present
in these theories and corresponds to the possible internal rotations
of the $D$ Majorana supercharges into each other. The supercharges
now transform as matrices under this twisted rotation group
and can hence be expanded on a basis of products of gamma matrices
\beq q=QI+Q_\mu\gamma^\mu+Q_{\mu\nu}\gamma^{\mu\nu}+\ldots\eeq
where the coefficients are the twisted supercharges. 
The original SUSY algebra 
\beq \{q_\alpha^I,\overline{q}_\beta^J\}=2\delta^{IJ}
\gamma^\mu_{\alpha\beta}P^\mu\eeq
now implies a twisted algebra
\beq
\{q,q\}_{\alpha\beta}=4\gamma^\mu_{\alpha\beta}P^\mu\\
\eeq
which naturally includes the nilpotent scalar supercharge $Q$.
Actually the twisted algebra also implies that the momentum is now $Q$-exact
\beq \{Q,Q_\mu\}=P_\mu \eeq
This property makes it plausible that the entire energy-momentum
tensor and hence action  of the theory may be $Q$-exact. This is
in agreement with the BRST form we have exhibited in the previous
examples.
Finally we should point out that we can match the
four supercharges of original SUSY theory by taking the twisted
supercharges to be real. This will imply a reality condition  on the
supercharge matrix
\beq q^\dagger=Cq^TC^{-1}\eeq
where $C$ will be the charge conjugation matrix.
If the supercharges form a matrix so do the fermions
which hence can be written in terms of anticommuting, antisymmetric tensor
fields $\eta$,$\psi_\mu$,$\chi_{\mu\nu}$ etc. 
We can abstract these p-form components and consider the fermions
as represented by a real K\"{a}hler-Dirac field 
\cite{2dpaper,4dpaper,noboru2,noboru4}
The original Dirac equation can then be shown to be
equivalent to the tensor \DK equation
\beq (d-d^\dagger)\Psi=0\eeq
where $d$ and $d^\dagger$ are the usual exterior derivative and its adjoint.
This corresponds to a fermion kinetic term (or \DK action)
\beq S_F=\Psi^\dagger .(d-d^\dagger)\Psi\eeq
This equivalence of the $D$-dimensional \DK equation to the Dirac equation 
for $D$ fermions remains
true when the model is gauged. In the continuum
this equivalence has been remarked on many times -- see for example
\cite{sol}. Hamiltonian lattice theories using \DK fermions
were first proposed in \cite{schwim}.

In this way we have exhibited the general change of variables implied
by the twist, exhibited the nilpotent
supercharge explicitly, and shown that such actions can be
written in a $Q$-exact form. What remains is to write down the
nilpotent symmetry and gauge fermion in a gauge theory and then
describe the prescription used to discretize the theory.
\section{Twisted ${\cal N}=2$ SYM in two dimensions}
The two dimensional \DK field representing the fermions contains 4 
grassman components 
$(\eta ,\psi_\mu,\chi_{12})$. Their corresponding commuting $Q$-superpartners
are labeled $(\phib, A_\mu, B_{12})$. All these fields take
values in the
adjoint of a gauge group with continuum twisted action
\cite{sugino,2dpaper}
\begin{eqnarray}
S&=&\beta Q{\rm Tr}\int
d^2x\left(
\frac{1}{4}\eta[\phi,\phib]+2\chi_{12}F_{12}+
\right.\nonumber\\
&+&\left.\chi_{12}B_{12}+\psi_\mu D_\mu \phib\right)\end{eqnarray}
where the scalar symmetry is given by
\begin{eqnarray}
QA_\mu&=&\psi_\mu\nonumber\\
Q\psi_\mu&=&-D_\mu\phi\nonumber\\
Q\phi&=&0\nonumber\\
Q\chi_{12}&=&B_{12}\nonumber\\
QB_{12}&=&[\phi,\chi_{12}]\nonumber\\
Q\phib&=&\eta\nonumber\\
Q\eta&=&[\phi,\phib]
\end{eqnarray}
The square of $Q$ is now an infinitessimal gauge transformation given by
the field $\phi$.
Carrying out the $Q$-variation and integrating out $B_{12}$ yields
the on-shell action
\begin{eqnarray}
S&=&\beta {\rm Tr}\int d^2x\left(
\frac{1}{4}[\phi,\phib]^2-\frac{1}{4}\eta [\phi,\eta]-F_{12}^2\right.\nonumber\\
&-&D_\mu \phi D_\mu \phib-\chi_{12}[\phi,\chi_{12}]\nonumber\\
&-&2\chi_{12}\left(D_1\psi_2-D_2\psi_1\right)-2\psi_\mu D_\mu\eta/2\nonumber\\
&+&\left.\psi_\mu [\phib,\psi_\mu]\right)
\end{eqnarray}
Notice that the scalar plus gauge part is positive definite 
along the contour $\overline{\phi}^a=\left(\phi^a\right)^*$ 
(we use AH group generators) and clearly corresponds to
the bosonic sector of 2D ${\cal N}=2$ super Yang-Mills. The fermionic
piece is nothing more than the \DK action described earlier.
\section{Twisted ${\cal N}=4$ SYM in four dimensions}
The four dimensional \DK field representing the fermions contains the
sixteen grassman components
$(\eta,\psi_\mu,\chi_{\mu\nu},\theta_{\mu\nu\lambda},\kappa_{1234})$
\cite{noboru4,4dpaper}.
Their corresponding $Q$-superpartners are labeled
$(\phib,A_\mu,B_{\mu\nu},W_{\mu\nu\lambda},C_{1234})$. The 
corresponding $Q$-transformations are generalizations of the
two dimensional ones:
\begin{eqnarray}
Q\phib&=&\eta\;\;Q\eta=[\phi,\phib]\nonumber\\
QA_\mu&=&\psi_\mu\;\;Q\psi_\mu=-D_\mu\phi\nonumber\\
QB_{\mu\nu}&=&[\phi,\chi_{\mu\nu}]\;\;Q\chi_{\mu\nu}=B_{\mu\nu}\nonumber\\
QW_{\mu\nu\lambda}&=&\theta_{\mu\nu\lambda}\;\;Q\theta_{\mu\nu\lambda}=
[\phi,W_{\mu\nu\lambda}]\nonumber\\
QC_{\mu\nu\lambda\rho}&=&[\phi,\kappa_{\mu\nu\lambda\rho}]\;\;Q\kappa_{\mu\nu\lambda\rho}=
C_{\mu\nu\lambda\rho}\nonumber\\
Q\phi&=&0
\label{Q}
\end{eqnarray}
Clearly $B$ and $C$ will be multiplier fields which are integrated
out to yield the on-shell supersymmetric action. The four fields of
$W$, together with $\phi$ and $\phib$ correspond to the usual six scalars
of ${\cal N}=4$ super Yang-Mills.
The appropriate gauge fermion is given by
$S=\beta Q\Lambda$ with
\begin{eqnarray}
\Lambda&=&\int d^4x {\rm Tr}\left[
\chi_{\mu\nu}\left(F_{\mu\nu}+\frac{1}{2}B_{\mu\nu}-
\frac{1}{2}[W_{\mu\lambda\rho},W_{\nu\lambda\rho}]\right.\right.\nonumber\\
&+&\left.\left.D_\lambda W_{\lambda\mu\nu}\right)\right.\nonumber\\
&+&\left.\psi_\mu D_\mu\phib+\frac{1}{4}\eta[\phi,\phib]+
\frac{1}{3!}\theta_{\mu\nu\lambda}[W_{\mu\nu\lambda},\phib]\right.\nonumber\\
&+&\left.\frac{1}{4!}\kappa_{\mu\nu\lambda\rho}\left(
\sqrt{2}D_{\left[\mu\right.}W_{\left.\nu\lambda\rho\right]}+
\frac{1}{2}C_{\mu\nu\lambda\rho}\right)
\right]
\label{qact}
\end{eqnarray}
Carrying out the $Q$-variation and subsequently
integrating out $B_{\mu\nu}$ and
$C_{\mu\nu\lambda\rho}$ leads to
\beq S=\beta\left(S_B+S_F+S_Y\right)\eeq
where
\begin{eqnarray}
S_F&=&\int d^4 x {\rm Tr}\left[
-\chi_{\mu\nu}D_{\left[\mu\right.}\psi_{\left.\nu\right]}
-\chi_{\mu\nu}D_\lambda\theta_{\lambda\mu\nu}\right.\\
&-&\left.\eta D_\mu\psi_\mu-
\frac{\sqrt{2}}{4!}\kappa_{\mu\nu\lambda\rho}D_{\left[\mu\right.}
\theta_{\left.\nu\lambda\rho\right]}
\right]
\end{eqnarray}
\begin{eqnarray}
S_B&=&\int d^4 x {\rm Tr}\left[
-\frac{1}{2}\left(
\left(F_{\mu\nu}-
\frac{1}{2}[W_{\mu\lambda\rho},W_{\nu\lambda\rho}]\right)^2\right.\right.\\
&+&\left.\left(D_\lambda W_{\lambda\mu\nu}\right)^2+
\frac{2}{4!}\left(D_{\left[\mu\right.}
W_{\left.\nu\lambda\rho\right]}\right)^2
\right)\nonumber\\
&-&\left.D_\mu\phi D_\mu\phib+\frac{1}{4}[\phi,\phib]^2-
\frac{1}{3!}[\phi,W_{\mu\nu\lambda}][\phib,W_{\mu\nu\lambda}]
\right]
\end{eqnarray}
We omit the Yukawas for simplicity. Again, a simple rescaling of the
fields renders the \DK nature of the fermionic action manifest while
the bosonic sector is nothing more than the Marcus twist of ${\cal N}=4$
super Yang-Mills after replacing the $W$-field by its dual
\cite{marcus}. Another
lattice formulation of ${\cal N}=4$ super Yang-Mills obtained from
the orbifold method was recently written down by Kaplan and Unsal \cite{kapn=4}.
\section{Lattice prescription} 
These twisted gauge actions may be discretized in a natural way.
We place 0-forms on sites, 1-forms on links, 2 forms on plaquettes etc.
For each orientation of the underlying p-cube we associate a field $f$
and its complex conjugate $f^\dagger$. Notice this complexification
doubles the degrees of freedom in the lattice theory with respect
to its continuum cousin. Furthermore, we choose the lattice fields to
have the following gauge transformation properties
\beq
f_{\mu_1\ldots\mu_p}(x)\to G(x)f_{\mu_1\ldots \mu_p}(x)G^{-1}(x+e_{\mu_1\ldots\mu_p})\eeq
where the vector $e_{\mu_1\ldots\mu_p}=\sum_{j=1}^p\mu_j$. 
A covariant forward difference operator is also defined by \cite{adjoint}
\beq
D^+_\mu f_{\mu_1\ldots\mu_p}(x)=
U_\mu(x)f_{\mu_1\ldots\mu_p}(x+\mu)-
f_{\mu_1\ldots\mu_p}(x)U_\mu(x+e_{\mu_1\ldots\mu_p})\eeq
and its adjoint a covariant difference operator via
\beq
D^-_\mu f_{\mu_1\ldots\mu_p}(x)=
f_{\mu_1\ldots\mu_p}(x)U^\dagger_\mu(x+e_{\mu_1\ldots\mu_p}-\mu)-
U^\dagger_\mu(x-\mu)f_{\mu_1\ldots\mu_p}(x-\mu)\eeq
These reduce to continuum derivatives as $a\to 0$ and ensure that
derivatives transform correctly under gauge transformations.
We also replace the continuum vector potential $A_\mu$ by the Wilson
gauge link $U_\mu$ which is to be treated as a non-unitary matrix at this
stage of the construction.

It was proved in \cite{rabin} that theories formulated in these geometrical
terms can be discretized without encountering spectrum doubling  if
\begin{eqnarray}
\partial_\mu&\to& D^+ {\rm \; if\; acts\; like\; d}\nonumber\\
\partial_\mu&\to& D^- {\rm \;if\; acts\; like\; d^\dagger}
\end{eqnarray}
We also use the following definition of the Yang-Mills field strength
\beq
F_{\mu\nu}(x)=D^+_\mu U_\nu(x)\to F^{\rm cont}_{\mu\nu}\;{\rm as}\;a\to 0\eeq
Using these ingredients we can straightforwardly construct the
lattice theory for both
continuum twisted theories. The $Q$-transformations are almost unchanged --
the only subtlety is that the explicit covariant derivative appearing
on the right hand side of the variation of $\psi_\mu$ must be a {\it
forward} difference operator and various commutators are point split
in such a way as to transform correctly under lattice gauge
transformations. We refer the reader to \cite{2dpaper,4dpaper} for
details. The discretization of the gauge fermion is straightforward;
we replace any term of the form
\beq\int A_{\mu_1\ldots\mu_p} B_{\mu_1\ldots\mu_p}\eeq
by the lattice expression
\beq\sum A^\dagger_{\mu_1\ldots\mu_p} B_{\mu_1\ldots\mu_p}+{\rm h.c}\eeq

This lattice theory is formulated in terms of complex fields. In \cite{4dpaper}
we give arguments that the theory can be truncated to the real line and
the twisted supersymmetric Ward identities recovered in the
continuum limit without additional fine
tuning.
\section{Conclusions}
It is possible to find formulations of a variety of supersymmetric theory which
can be written in the language of differential forms and exterior
derivatives. The fermion content of such theories may be embedded
in one (or more) \DK fields. Such a theory has a $Q$-exact action
and a scalar nilpotent supercharge. The latter generates a fermionic
symmetry which may be implemented exactly on the lattice. We have
illustrated this with examples drawn from quantum mechanics, 
two dimensional sigma models and Yang-Mills theories.
The use of \DK fermions evades the standard doubling problems and allows
local, $Q$-symmetric lattice actions to be written
down. In the case of gauge theories the requirements of gauge
invariance force a complexification of the degrees of freedom. 
Significant numerical work has already been
done in the non-gauge models and is currently starting in the
Yang-Mills case. We hope that these studies, complemented by
perturbative calculations will help establish these lattice theories
as good non-perturbative regulators of the corresponding
continuum theories. If this proves correct, then they may be used
to explore the strong coupling physics of models such as
large N Yang-Mills, which would give us a new non-perturbative handle
on various string and supergravity theories. 
\section*{Acknowledgments}
The
author would like to thank the Summer Institute 2005 at Fuji-Yoshida
for providing a stimulating atmosphere and the Yukawa Institute for
Theoretical Physics at Kyoto for financial support.  Additional
support for this work
was provided by DOE grant DE-FG02-85ER40237.

\end{document}